\documentclass[conference]{IEEEtran}
\IEEEoverridecommandlockouts
\usepackage{booktabs} 
\usepackage{graphicx}
\usepackage{amsfonts}
\usepackage{dsfont}
\usepackage{graphicx, epsfig,amsmath,amssymb,latexsym,graphics,float}
\usepackage{setspace,subfig}
\usepackage{citesort}
\usepackage{float}
\usepackage[ruled,vlined,linesnumbered]{algorithm2e}
\usepackage{booktabs}
\usepackage{lipsum}
\usepackage{multicol}
\usepackage{url}
\usepackage{multirow}
\usepackage[utf8]{inputenc}
\usepackage{graphicx}
\usepackage{amsmath}
\usepackage{booktabs}
\usepackage{algorithmic}
\begin{document}

\title{Temporal Data Fusion at the Edge\\
\thanks{$^{\star}$Correspondence author. This work was partly supported by National Natural Science Foundation of China (Nos. 61571238, 61906099), Scientific Research Foundation of Nanjing University of Posts and Telecommunications (No. NY218072) and a research fund from Yancheng Big Data Research Institute.
}
}
\author{
\IEEEauthorblockN{Linfu Yang}
\IEEEauthorblockA{\textit{School of Computer Science}\\
\textit{Nanjing University of Posts and Telecommunications}\\
Nanjing, China\\
1017041212@njupt.edu.cn}
\and
\IEEEauthorblockN{Bin Liu$^{\star}$}
\IEEEauthorblockA{\textit{School of Computer Science}\\
\textit{Jiangsu Key Lab of Big Data Security $\&$ Intelligent Processing} \\
\textit{Nanjing University of Posts and Telecommunications}\\
Nanjing, China\\
bins@ieee.org}
}

\maketitle

\begin{abstract}
As an enabler technique, data fusion has gained great attention in the context of Internet of things (IoT).
In traditional settings, data fusion is done at the cloud server. So the data to be fused should be transferred from
the sensor nodes to the cloud server before data fusion. Such an application mode of data fusion inherits disturbing
concerns from the cloud computing framework, e.g., privacy-leaking, large latency between data capture and computation,
excessive ingress bandwidth consumption. We take into account how to do temporal data fusion at the edge to bypass the above issues.
We present a Gaussian process based temporal data fusion (GPTDF) method targeted for the problem of sequential online prediction
at the edge. The GPTDF method fits the edge computing framework and thus inherits desirable properties from edge computing, such as
privacy-preserving, low latency between data capture and computation, and tiny bandwidth consumption. Through a real-data experiment using archived traffic datasets from the Caltrans Performance Measurement System (PeMS), we demonstrate that the application of GPTDF can provide more timely and accurate real-time predictions at the network edge.
\end{abstract}


\section{Introduction}
In recent years, the Internet of things (IoT) has become ubiquitous due to the advances in sensor and computing technologies and
commercial needs from manufacturing industries, smart farming, to autonomous
vehicles \cite{al2015internet,ahmed2016internet,bi2014internet,kamilaris2016agri,gerla2014internet}.
As a result of it, there is an exponential increase in the number of network nodes connected to the Internet,
which then generates an extreme amount of data that need to be stored and analyzed in a timely fashion.
As an enabler technique for data analysis, data fusion has recently gained great attention in the field
of IoT \cite{wang2016city,dautov2017distributed,dautov2017three,singh2014survey}.

Broadly speaking, data fusion refers to
the theory, techniques and tools applied for combining relevant information from multiple sources to provide better
decisions or actions than would not be possible if any of these data sources was used individually. Cloud computing is integrated
with IoT to handle the massive data \cite{botta2014integration,botta2016integration}. The cloud server can provide elastic
virtual resources management, storage capacity, and computation facility. Currently, most of the data fusion processing for
IoT is done on the cloud server.

The traditional cloud-based data fusion procedure requires that all data be transferred from the data sources to the cloud server
before data fusion. This leads to disturbing concerns, e.g., privacy-leaking, large latency between data capture and computation,
and excessive ingress bandwidth consumption.

\begin{figure}[!htb]
\centering
\includegraphics[width=3.2in,height=2in]{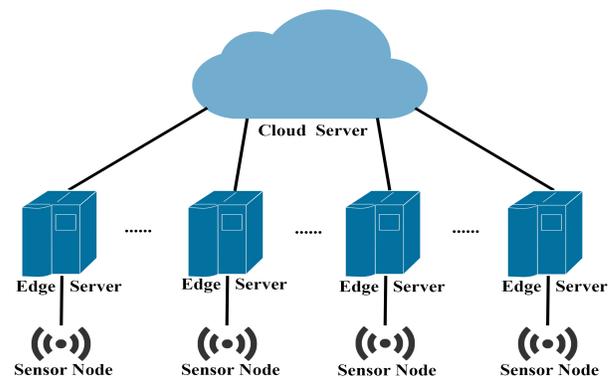}
\caption{A schematic diagram of edge computing}\label{fig:edge}
\end{figure}

Fog computing and edge computing have emerged as the new alternative paradigms of cloud
computing. They provide possibilities to process data near or at the data source rather than transferring
data to the cloud \cite{bonomi2012fog,shi2016edge}. For a schematic diagram of edge computing, see Fig.\ref{fig:edge}, which shows that, for each sensor node, there is at least one edge server dispatched close to it.
The edge server is responsible for processing the data generated at the sensor node. In contrast with cloud computing,
which suffers from the inherent speed of light latency, edge computing can enable real-time data processing with negligible
latency due to the close distance between the sensor node and the edge server. This is a desirable property for time-sensitive
applications like autonomous vehicles. Besides, by processing the data at the edge, the data privacy is preserved and the bandwidth for data relaying is saved.

In this paper, we consider how to do temporal data fusion at the network edge. The goal is to combine strengths of temporal data fusion
and edge computing. As coined in \cite{hsiao2005temporal}, temporal data fusion refers to the fusion of data or information
acquired over some time. Different from traditional fusion methods that only fuse sensor data at a point in time,
temporal fusion aims at inferring dynamic patterns of the system rather than just the system state at a point in time. For clarity,
here we take traffic flow prediction as one application instantiation of temporal data fusion. Specifically, we consider how to achieve more timely and accurate real-time predictions at a target edge node by borrowing knowledge from data analyzed at other edge nodes. The challenge to address the above issue lies in that, on one side, we would like to borrow as much related knowledge as possible to overcome the cold start problem when launching the prediction algorithm at the target edge node; meanwhile, on the other side, we hope to transmit as less data as possible among the network nodes to save communication bandwidth, reduce data processing latency and preserve data privacy.
We break the above dilemma through a novel algorithm design termed Gaussian process (GP) based temporal data fusion (GPTDF), the efficacy and accuracy of which is demonstrated using a real-data based experiment.

To summarize, the main contribution of this paper is two-fold. First, we propose the concept of temporal data fusion at the edge. To the best of our knowledge, this is the first paper that introduces this concept to the literature. Second, we propose a novel algorithm design, namely GPTDF, which works at the edge to provide sequential online prediction service.
The remainder of this paper is organized as follows. Section \ref{sec:gpts} briefly introduces the Gaussian process (GP) based approach for capturing the temporal feature from time-series data. Section \ref{sec:gptdf} presents the GPTDF method. Section \ref{sec:connection} summarizes the connections and differences between our GPTDF method and the relevant works in the literature. Section \ref{sec:experiment} provides experimental results on the application of GPTDF for sequential online traffic flow prediction. Finally, Section \ref{sec:conclusion} concludes the paper.
\section{Gaussian Process based Temporal Feature Capturing}\label{sec:gpts}
We treat a time-series, $\{t_i, y(t_i)\}, i=1, 2, \ldots, n$, as a random sample drawn from a Gaussian Process (GP).
Here $y(t_i)$ denotes the $i$th data point in the time-series, which is observed at time $t_i$.
A GP can be seen as a distribution over functions fully specified by a pair of a mean function and a covariance kernel function.
For more details on GP and its applications, readers are referred to \cite{williams2006gaussian}.

Here we use GP to model the mapping relationship from the time variable $t$ to the observation $y(t)$ as follows
\begin{equation}
y(t)=f(t), f \sim \mathcal{GP}\left(\mu, k_{\theta}\right),
\end{equation}
where $\mathcal{GP}\left(\mu, k_{\theta}\right)$ denotes a GP specified by the mean function $\mu(\cdot)$ and the covariance
kernel function $k_{\theta}(\cdot,\cdot)$ parameterized by $\theta$, $f$ is a random function drawn from this GP.
Given a set of input locations $\mathbf{t}=\{t_1,\ldots,t_i\}$,
the covariance elements associated with each pair of the input locations can be described by the covariance matrix
\begin{equation}
\mathbf{K}_{\theta}(\mathbf{t}, \mathbf{t})=\left(\begin{array}{cccc}{k_{\theta}\left(t_{1}, t_{1}\right)} &
{k_{\theta}\left(t_{1}, t_{2}\right)} & {\dots} & {k_{\theta}\left(t_{1}, t_{i}\right)} \\
{k_{\theta}\left(t_{2}, t_{1}\right)} & {k_{\theta}\left(t_{2}, t_{2}\right)} & {\dots} & {k_{\theta}\left(t_{2}, t_{i}\right)} \\
{\vdots} & {\vdots} & {\vdots} & {\vdots} \\ {k_{\theta}\left(t_{i}, t_{1}\right)} & {k_{\theta}\left(t_{i}, t_{2}\right)} &
{\dots} & {k_{\theta}\left(t_{i}, t_{i}\right)}\end{array}\right),
\end{equation}
where $k_{\theta}\left(t_{j}, t_{k}\right)$ denotes the covariance element between $t_j$ and $t_k$, $j, k\in\{1,\ldots,i\}^2$.
Then evaluations of $f$ at input locations covered in $\mathbf{t}$ can be taken as a draw from a multi-variate Gaussian distribution,
\begin{equation}
p(\mathbf{y}(\mathbf{t}))=\mathcal{N}(\boldsymbol{\mu}(\mathbf{t}), \mathbf{K}_{\theta}(\mathbf{t}, \mathbf{t})).
\end{equation}
Here $\mathbf{y}(\mathbf{t})=\left\{y_{1}, y_{2}, \ldots, y_{i}\right\}$ denotes dependent function values evaluated at
$t_{1}, t_{2}, \ldots, t_{i}$, respectively; $\boldsymbol{\mu}(\mathbf{t})$ denotes the mean vector that consists of mean function values, again evaluated at $t_{1}, t_{2}, \ldots, t_{i}$, respectively.
To take account of the observation noise, we can add a noise item $\eta$ as follows
\begin{equation}\label{eqn:obs_fun}
y(t)=f(t)+\eta,
\end{equation}
where $\eta$ is assumed to be Gaussian distributed, namely $\eta \sim \mathcal{N}\left(0, \sigma_n^{2}\right)$, where $\sigma_n^{2}$ denotes the variance. Then the form of the covariance matrix becomes
\begin{equation}\label{eqn:V}
\mathbf{V}_{\theta}(\mathbf{t}, \mathbf{t})=\mathbf{K}_{\theta}(\mathbf{t}, \mathbf{t})+\sigma_n^{2} \mathbf{I}
\end{equation}
where $\mathbf{I}$ denotes the identity matrix.

\begin{figure}[!htb]
\centering
\includegraphics[width=1.8in,height=1.1in]{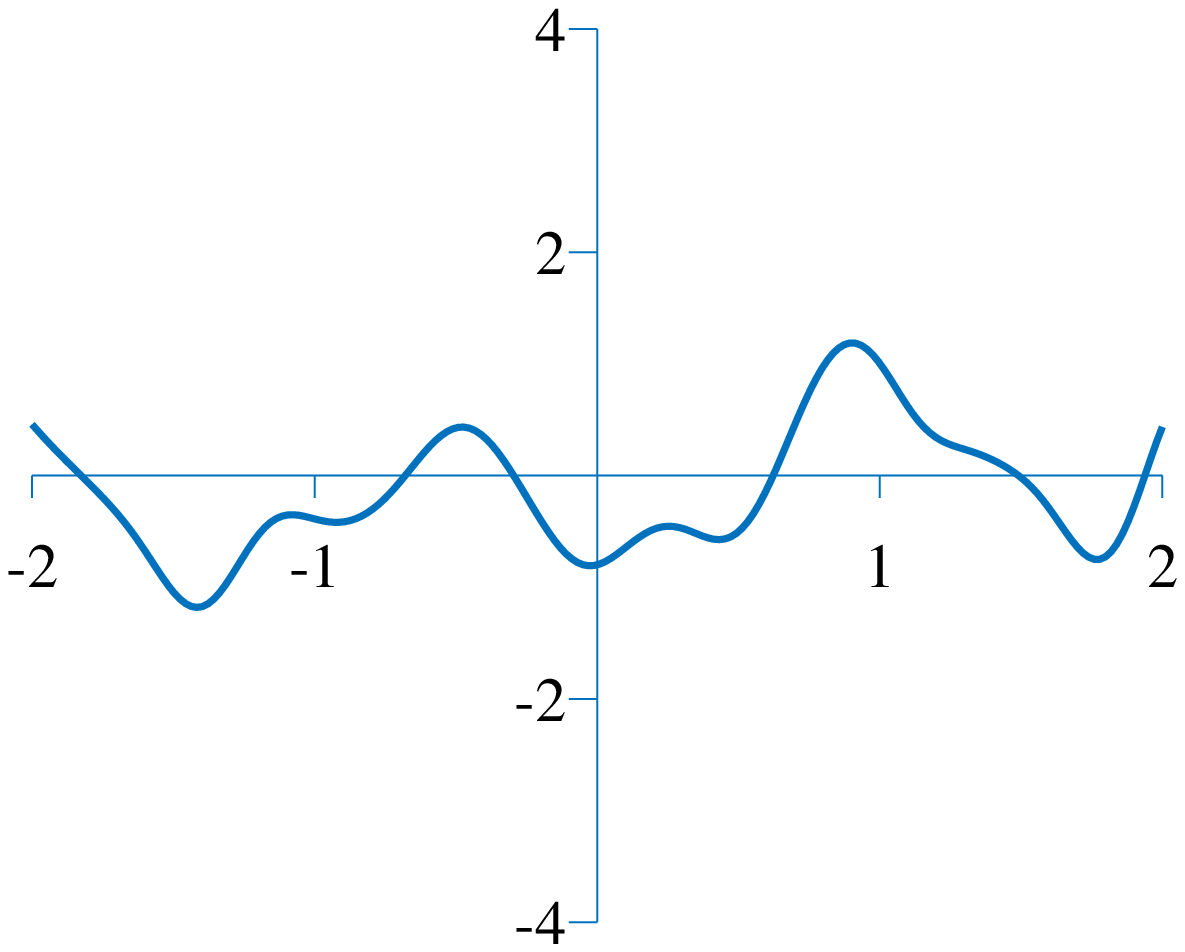}\includegraphics[width=1.8in,height=1.1in]{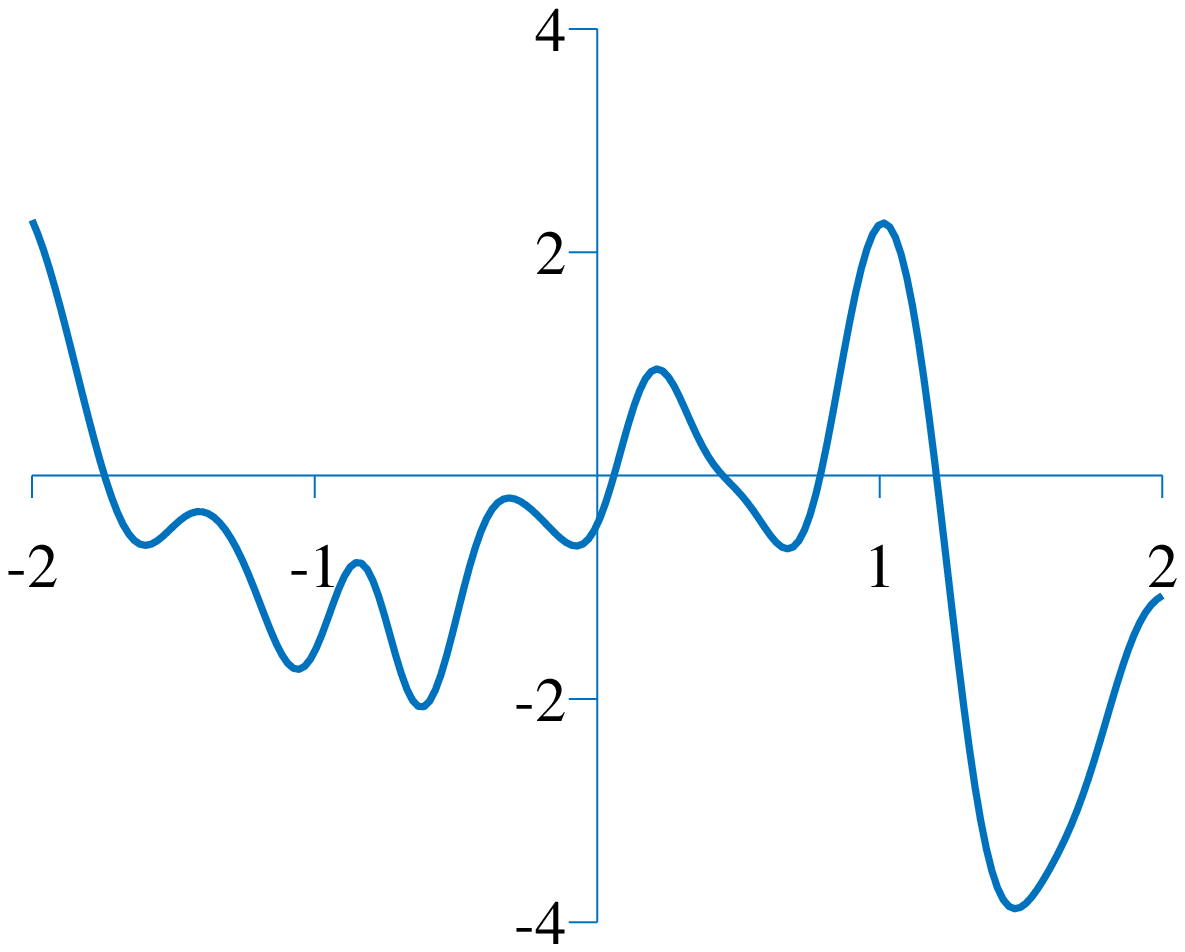}\\
\includegraphics[width=1.8in,height=1.1in]{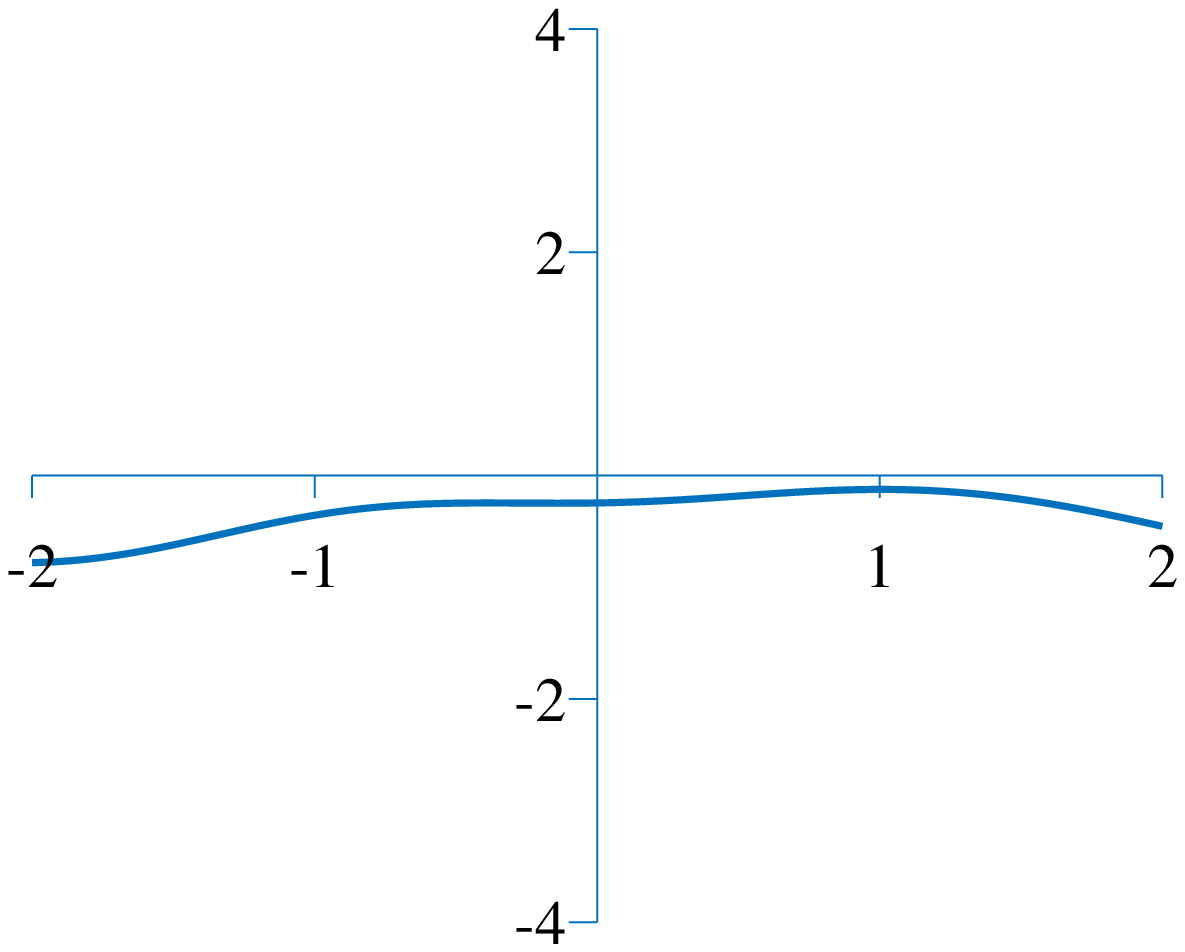}\includegraphics[width=1.8in,height=1.1in]{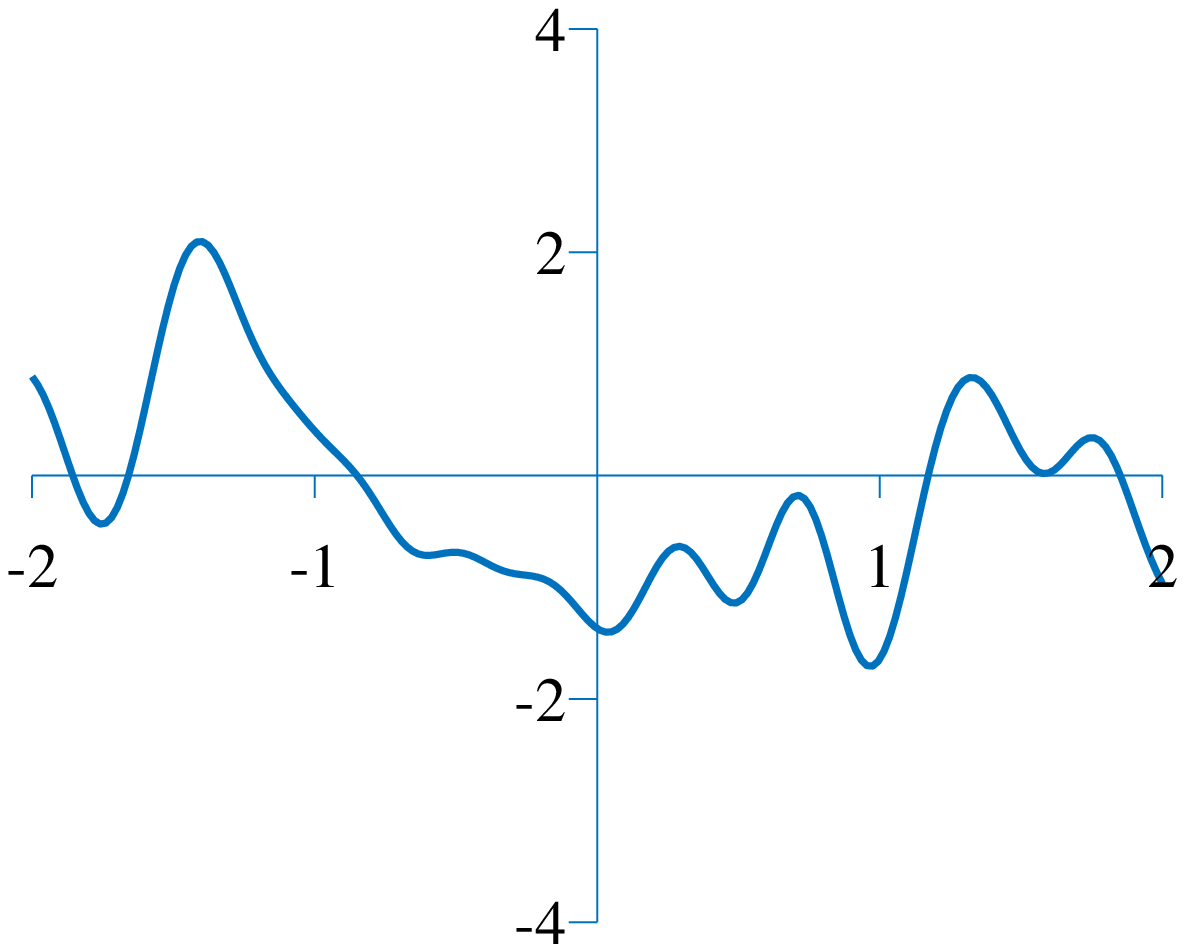}
\caption{Random function curves drawn from a GP that uses a zero-valued mean function and an SE type covariance kernel function.
Clockwise from top-left, the associated GP hyper-parameters $[h^2, \lambda, \sigma_n]$ take values $[0.5,\,0.2,\,0], [1.5,\,0.2,\,0], [0.5,\,0.2,\,2], [0.5,\,1,\,0]$, respectively.}\label{fig:gp_sample}
\end{figure}

The covariance kernel function can take different forms. For example, the squared exponential (SE) function, often adopted as the
covariance kernel function, is given by \cite{williams2006gaussian}
\begin{equation}\label{eq:se}
k_{\theta}\left(t_{i}, t_{j}\right)=h^{2} \exp \left[-\left(\frac{t_{i}-t_{j}}{\lambda}\right)^{2}\right].
\end{equation}
Its hyper-parameters $\theta\triangleq[h,\lambda]$ describe general properties of our function $f$ \cite{williams2006gaussian}.
Specifically, the parameter $h$ governs the output scale of $f$, $\lambda$ determines its input scale,
and thus its smoothness.
In Fig.\ref{fig:gp_sample}, we show four random functions sampled from the GP, each corresponding to a specific setting of hyper-parameter values.
We see that the GP hyper-parameters can provide a quantitative and succinct description for the associated time-series data.

Here we adopt the Matern $5/2$ kernel function \cite{williams2006gaussian}, given by 
\begin{equation}\label{eqn:matern}
k_{\theta}\left(x_{i}, x_{j}\right)=\sigma_{f}^{2}\left(1+\frac{\sqrt{5} r}{\sigma_{l}}+\frac{5 r^{2}}{3 \sigma_{l}^{2}}\right) \exp \left(-\frac{\sqrt{5} r}{\sigma_{l}}\right),
\end{equation}
where $r=\sqrt{\left(t_{i}-t_{j}\right)^{T}\left(t_{i}-t_{j}\right)}$ is the Euclidean distance between $t_{i}$ and $t_{j}$, $\theta\triangleq[\sigma_{f},\sigma_{l}]$. Now it is the parameter $\sigma_{f}$ that governs the output scale of our function, and $\sigma_{l}$ describes its smoothness. Given a GP approximation of the time-series $\{t_i, y(t_i)\}, i=1, 2, \ldots, n$, we use parameters $\epsilon\triangleq\{\sigma_{f},\sigma_{l},\sigma_n\}$ to describe the temporal structure of this time-series.
Given an observed dataset $\{\mathbf{t},\mathbf{y}\}$, we set the value of $\epsilon$ by maximizing the log marginal likelihood \cite{williams2006gaussian}:
\begin{eqnarray}\label{eqn:mlik}
\log p(\mathbf{y}|\mathbf{t})&=&-\frac{1}{2} \mathbf{y}^{\top}\left(\mathbf{V}_{\theta}(\mathbf{t}, \mathbf{t})\right)^{-1} \mathbf{y}\\\nonumber
& &-\frac{1}{2} \log \left|\mathbf{V}_{\theta}(\mathbf{t}, \mathbf{t})\right|-\frac{n}{2} \log 2 \pi.
\end{eqnarray}
For approaches to solve the above optimization problem, see \cite{rasmussen2010gaussian}.
As shown above, GP provides a way to capture temporal features from time-series data. GP also provides a way to do prediction.
Let consider predicting the data point $y_{*}$ that will be observed at a future time $t_{*}$ based on
an observed dataset $\{\mathbf{t},\mathbf{y}\}$.
According to the definition of GP, it can be derived that the distribution of $y_{*}$ conditional on $\{\mathbf{t},\mathbf{y}\}$ is
Gaussian with mean \cite{williams2006gaussian}
\begin{equation}\label{eqn:mean}
m_{*}=\mu\left(t_{*}\right)+\mathbf{K}_{\theta}\left(t_{*}, \mathbf{t}\right) \mathbf{V}_{\theta}(\mathbf{t}, \mathbf{t})^{-1}(\mathbf{y}-\boldsymbol{\mu}(\mathbf{t}))
\end{equation}
and variance
\begin{equation}\label{eqn:variance}
\sigma_{*}^{2}=k_{\theta}\left(t_{*}, t_{*}\right)-\mathbf{K}_{\theta}\left(t_{*}, \mathbf{t}\right) \mathbf{V}_{\theta}(\mathbf{t}, \mathbf{t})^{-1} \mathbf{K}_{\theta}\left(\mathbf{t}, t_{*}\right).
\end{equation}
\section{The Proposed GPTDF Method for Sequential Online Prediction at the Edge}\label{sec:gptdf}
The GPTDF algorithm is targeted for the problem of sequential online prediction at the edge.
We take this problem as one application instantiation of the concept of temporal data fusion at the edge.

To do predictions with a learning algorithm, it is required that enough labeled data be collected beforehand for
training a prediction model. For the GP model being used here, its hyper-parameters need to be determined in the training procedure before making predictions. This leads to a dilemma termed cold starting. That says, if a new edge server is added into the system, then the prediction can not be performed immediately, since there is no data stored there for training the prediction model.

We propose the GPTDF method to solve the above dilemma (See Algorithm \ref{alg:GPTDF} for a pseudo-code to implement it). First, let the target edge server query the cloud server about the
temporal features of historical datasets that have been processed at other edge servers. Then the cloud server sends back related temporal
feature data to the target edge server. As described in Section \ref{sec:gpts}, the temporal feature data corresponding
a time-series dataset only consists of three parameters, namely $\sigma_f, \sigma_l$, and $\sigma_n$.
\begin{algorithm}[tb]
\caption{The GPTDF method for sequential online prediction at the target edge server}
\label{alg:GPTDF}
\begin{algorithmic}[1] 
\STATE The target edge server sends queries to the cloud server;
\STATE The cloud server sends back to the target edge server $\{\sigma_{f,j}, \sigma_{l,j}, \sigma_{n,j}\}_{j=1}^M$;
\FOR{$i=1, \ldots, n-1$} \FOR{$j=1,\ldots,M$}
\STATE Calculate $m_{j,i+1}, \sigma_{j,i+1}^2$ based on the $j$th GP model whose hyper-parameters are $\sigma_{f,j}, \sigma_{l,j}, \sigma_{n,j}$. See (\ref{eqn:mean})-(\ref{eqn:variance}) for equations involved for the above calculation;
\ENDFOR
\STATE Calculate $\hat{\omega}_{j,i+1}$ with (\ref{eqn:pred_weight}), $j=1, \ldots, M$;
\STATE Calculate $\omega_{j,i+1}$ with (\ref{eqn:weight}), $j=1, \ldots, M$;
\STATE Calculate $m_{t+1}, \sigma_{t+1}^2$ using (\ref{eqn:mean_var})-(15);
\STATE Output $m_{t+1}, \sigma_{t+1}^2$.
\ENDFOR
\end{algorithmic}
\end{algorithm}
Suppose that, after the query, the cloud server sends back to the target edge
server $M$ groups of hyper-parameter values, $\{\sigma_{f,j}, \sigma_{l,j}, \sigma_{n,j}\}_{j=1}^M$,
each group standing for a report made by another edge server to the cloud server.
Now there are $M$ candidate GP models that can be used for sequential online prediction at the target edge server,
each model being characterized by a group of hyper-parameter values, namely, $\{\sigma_{f,j}, \sigma_{l,j}, \sigma_{n,j}\}$, $j\in\{1, \cdots, M\}$.

We use a weighted mixture of these models to capture the non-stationary temporal structure of the data
that will be observed at the target edge server. For implementing a data-driven automatic tuning of the model weights,
we resort to the dynamic model averaging (DMA) technique \cite{dai2016robust,liu2011instantaneous,liu2017robust,liu2019sequential}.
Suppose that, at time $t_i$, the weight of the model $\mathcal{M}_j$ is $\omega_{j,i}>0$, $j\in\{1,\ldots,M\}$, $\sum_{j=1}^M\omega_{j,i}=1$. Then the predictive weights of the models at time $t_{i+1}$ are defined to be
\begin{equation}\label{eqn:pred_weight}
\hat{\omega}_{j,i+1}=\frac{\omega_{j,i}^{\alpha}}{\sum_{k=1}^{M}\omega_{k,i}^{\alpha}}, j=1,\ldots,M,
\end{equation}
where $0<\alpha<1$ is termed the forgetting parameter. Upon the arrival of the observation $y(t_{i+1})$, the model weights are updated according to Bayesian formalism as follows
\begin{equation}\label{eqn:weight}
\omega_{j,i+1}=\frac{\hat{\omega}_{j,i+1} p\left(y(t_{i+1})|\mathcal{M}_j\right)}{\sum_{k=1}^{M}\hat{\omega}_{k,i+1} p\left(y(t_{i+1}) |\mathcal{M}_k\right)}, j=1,\ldots,M,
\end{equation}
where $p\left(y(t_{i+1})|\mathcal{M}_j \right)$ denotes the likelihood of the hypothesis $\mathcal{M}_j$ given $y(t_{i+1})$, $j=1, \ldots, M$.

To combine predictions provided by $\mathcal{M}_1, \ldots, \mathcal{M}_M$ to yield a fused prediction, we resort to the weighted version of the product of experts (PoE) model. Denote the predictive distribution of $y(t_{i+1})$ corresponding to $\mathcal{M}_j$ as $p_j(y(t_{i+1})|y(t_{i-\tau+1}),\ldots, y(t_i))$ (or $p_j(y(t_{i+1}))$ for short), where $\tau$ denotes the length of the time window. Then the fused predictive distribution of $y(t_{i+1})$ is defined to be
\begin{equation}\label{eqn:generalized_poe}
p(y(t_{i+1}))\propto \Pi_{j=1}^M \left[p_j(y(t_{i+1}))\right]^{\hat{\omega}_{j,i+1}}.
\end{equation}
Since $p_j(y(t_{i+1}))$, $j=1, \ldots, M$, are Gaussian, $p(y(t_{i+1}))$ calculated with (\ref{eqn:generalized_poe}) is still Gaussian, with its mean and variance given by \cite{cao2014generalized}
\begin{eqnarray}\label{eqn:mean_var}
m_{i+1} &=& \frac{\sum_{j=1}^M(m_{j,i+1}\hat{\omega}_{j,i+1}P_{j})}{\sum_{j=1}^M(\hat{\omega}_{j,i+1}P_{j})}, \\
\sigma_{i+1}^{2}&=&1/\sum_{j=1}^M(\hat{\omega}_{j,i+1}P_{j}),
\end{eqnarray}
where $P_j=\left(\sigma_{j,i+1}^{2}\right)^{-1}$, $m_{j,i+1}$ and $\sigma_{j,i+1}^{2}$ denote the mean and variance associated
with $p_j(y(t_{i+1}))$, respectively. The mean $m_{i+1}$ is taken as the prediction of $y(t_{i+1})$ made at time $t_i$.
A confidence interval associated with this prediction is also available.
For example, a 99.75\% confidence interval is shown to be $[m_{i+1}-3\sigma_{i+1}, m_{i+1}+3\sigma_{i+1}]$.

In GPTDF as shown in Algorithm \ref{alg:GPTDF}, the forgetting parameter $\alpha$ is initialized at 0.9, and  $\omega_{j,1}$ is set at $1/M$, for $j=1, \ldots, M$.
\subsection{Algorithm Analysis}\label{sec:analysis}
In GPTDF, for a dataset, only its temporal features $\sigma_{f}, \sigma_{l}, \sigma_{n}$ are transferred between the cloud server and the edge servers. Compared with the raw dataset, the size of the temporal feature data is much compressed. Therefore, transferring the feature data between an edge server and the cloud server may only consume negligible bandwidth and take little time. In addition, since only temporal features are transferred between the cloud server and the edge servers, the raw data are invisible for the cloud server and all edge servers except the one that is connected to the data source. That says the data privacy is preserved.
\section{Connections to Existent Works}\label{sec:connection}
The GPTDF method presented here is related with other GP model based time-series analysis methods in e.g., \cite{roberts2013gaussian,osborne2012real,liu2019sequential}, among which the instant temporal structure learning (INTEL) algorithm of \cite{liu2019sequential} is of most relevance. Both GPTDF and INTEL use multiple GP models. The crucial difference between them lies in that, in the former, each GP model is associated with one specific edge server together with one time-series data that have been analyzed there. That says, for each GP model involved in GPTDF, there is a unique training dataset associated with it. For the INTEL algorithm, except a template model itself, all the other GP models are built based on the template model. They are variants of the template model and there is no training dataset associated with any of them. Conceptually speaking, GPTDF provides a way to fuse different temporal datasets collected from different sensor nodes, while INTEL provides a way to make use of prior knowledge in processing one single temporal dataset.

Our temporal data fusion method proposed here is also relevant with existent data fusion methods developed for IoT applications in e.g., \cite{wang2016city,dautov2017distributed,dautov2017three}. The biggest difference between our approach and these existing methods is that our approach performs the fusion of temporal data and is run at the edge server, while most of these existent methods are run at the cloud server and do the fusion of non-temporal data.
\section{Experiment}\label{sec:experiment}
In this section, we focus on an edge computing application scenario, namely real-time traffic flow prediction at the edge.
We seek to experimentally validate that the proposed GPTDF method can provide more timely and accurate predictions at the edge.

In our experiment, we used archived traffic datasets from the Caltrans Performance Measurement System (PeMS) \cite{chen2001freeway}.
These data are collected in real-time from over 39,000 individual sensors, which span the freeway system across all major metropolitan
areas of the State of California. We selected 19 segments of time-series from the PeMS dataset, 18 of which are treated as historical
datasets that have been stored at 18 edge servers, respectively, and the other one is treated as the dataset that is observed and
processed at the target edge server. For each dataset, an identical pre-processing operation is used to do data normalization.
The normalized dataset has mean zero and standard error one.
\begin{table}[!t]\centering\small
\caption{Prediction performance comparison}
\begin{tabular}{c|c|c|c|c}
\hline %
  & NLL & MAE & MSE & Delay \\\hline
GPTDF-All & 0.2839  & 0.2472 & 0.1041 & 0\\\hline
GPTDF-I & 0.1591 & 0.2081 & 0.0777 & 0 \\\hline
GPTDF-II & 3.4316 & 0.3098 & 0.1673 & 0 \\\hline
GP-I ($N=50$) & 5.7005 & 0.2954 & 0.1652 & 50 \\\hline
GP-II ($N=100$)& 1.8171 & 0.2398 & 0.1078 & 100 \\\hline
GP-III ($N=150$)& 0.1560 & 0.2063 & 0.0788 & 150 \\\hline
\end{tabular}
\label{Table:compare}
\end{table}
\begin{figure}[!htb]
\centering
\includegraphics[width=1.8in,height=0.8in]{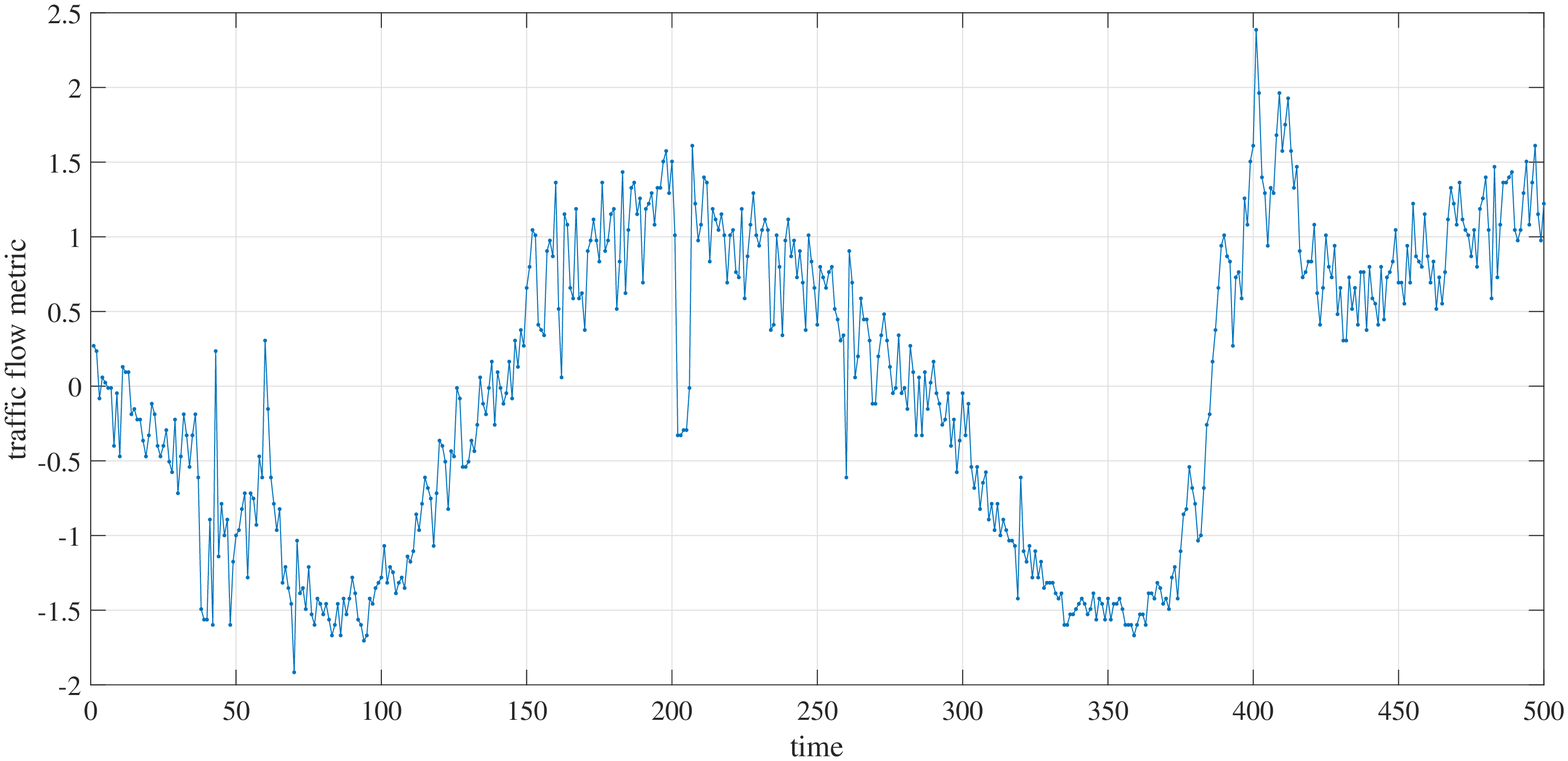}\includegraphics[width=1.8in,height=0.8in]{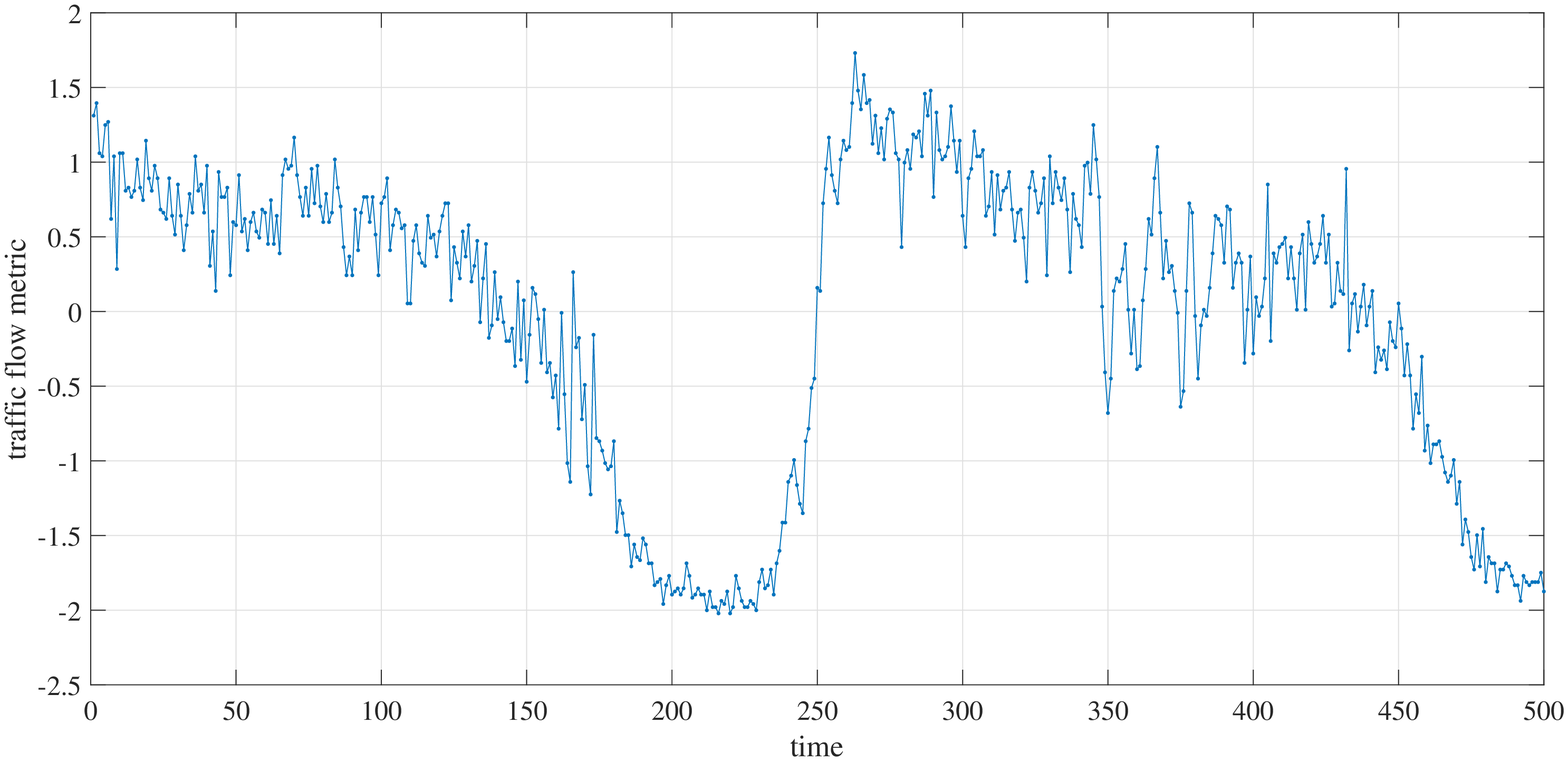}\\
\includegraphics[width=1.8in,height=0.8in]{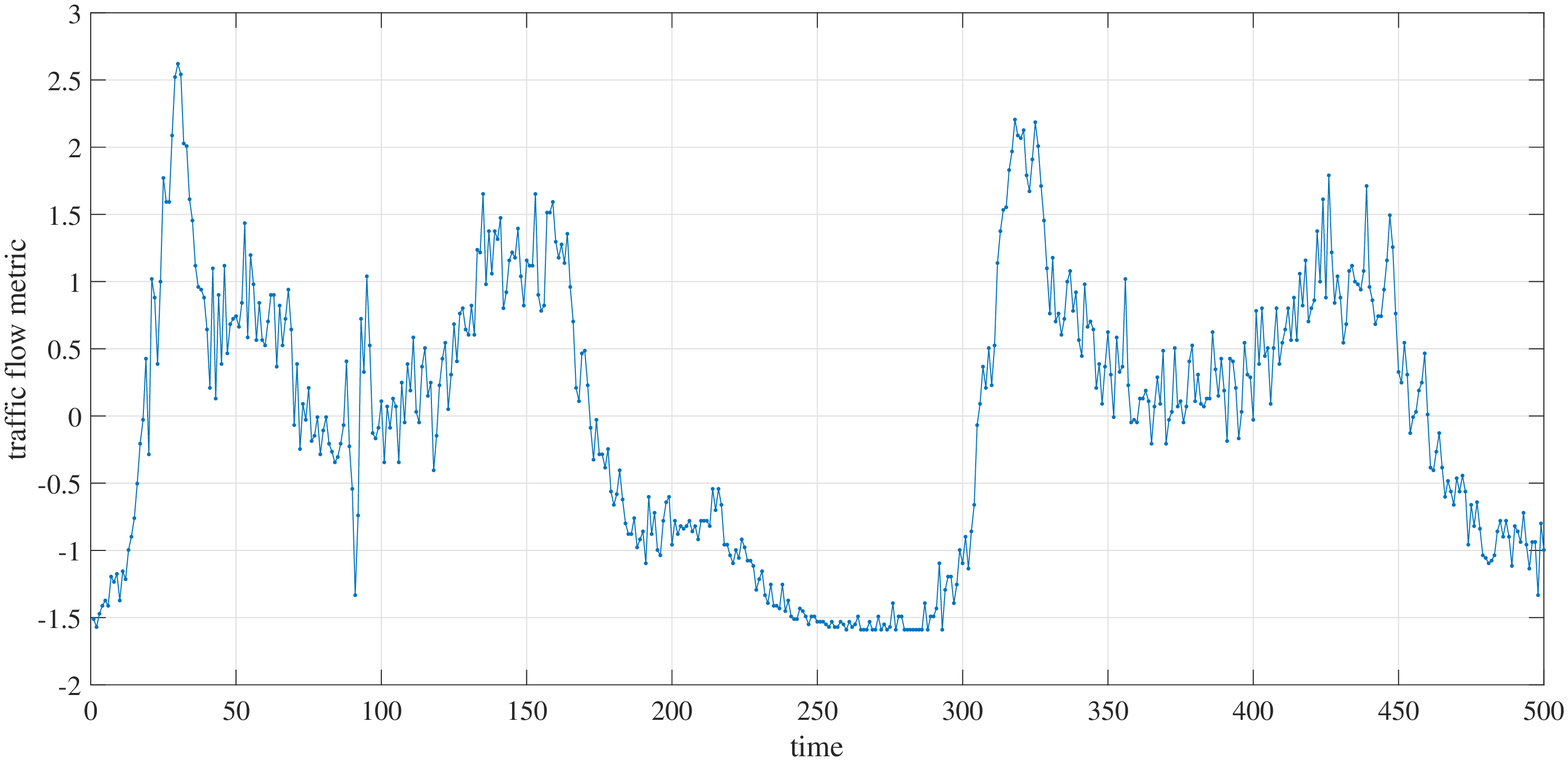}\includegraphics[width=1.8in,height=0.8in]{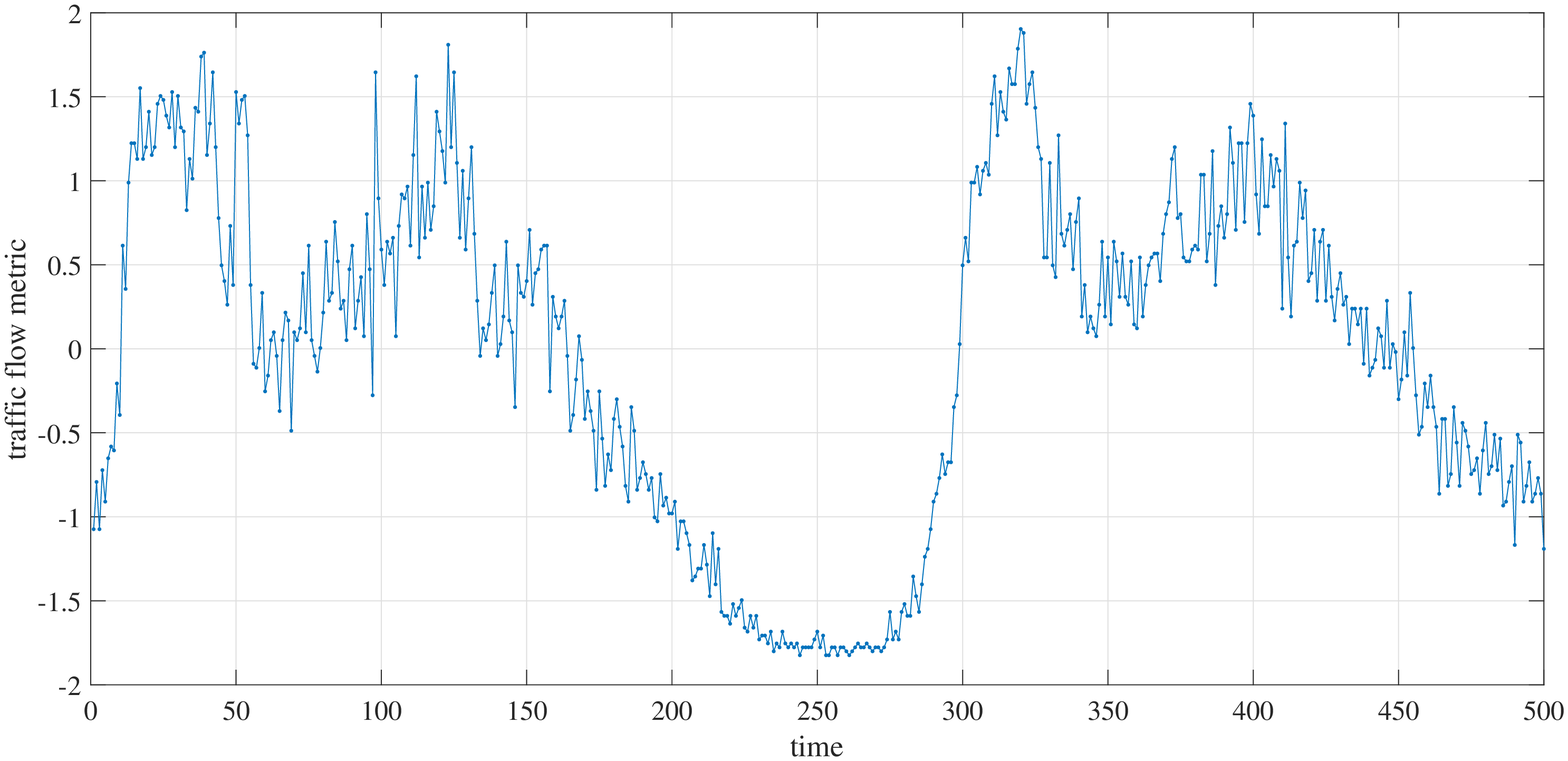}
\caption{Four segments of time-series from the PeMS dataset \cite{chen2001freeway}. They are treated as historical datasets that have been stored at 4 non-target edge servers. They are used by GPTDF-I for temporal data fusion.}\label{fig:dataset1}
\end{figure}

\begin{figure}[!htb]
\centering
\includegraphics[width=1.8in,height=0.8in]{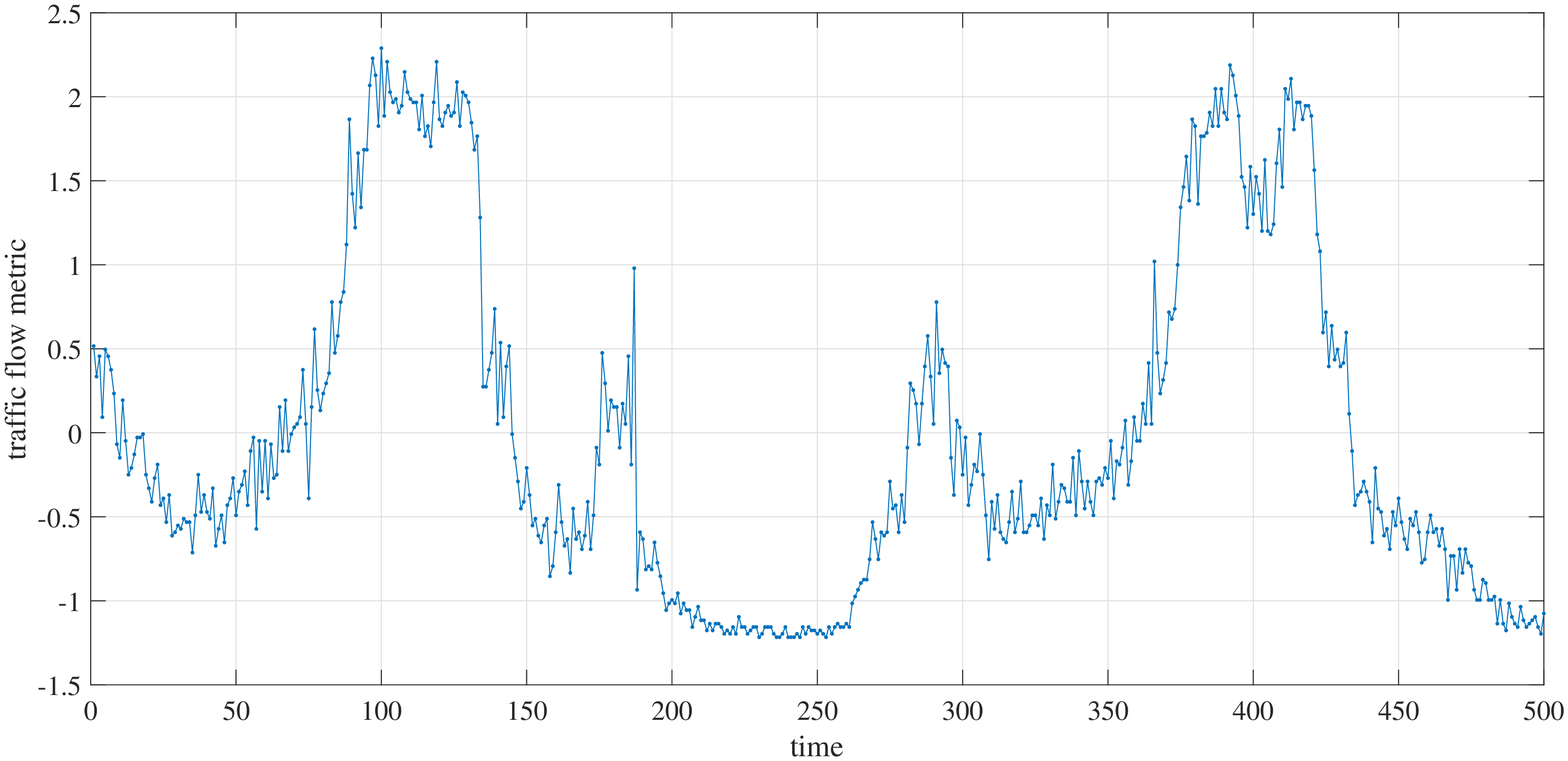}\includegraphics[width=1.8in,height=0.8in]{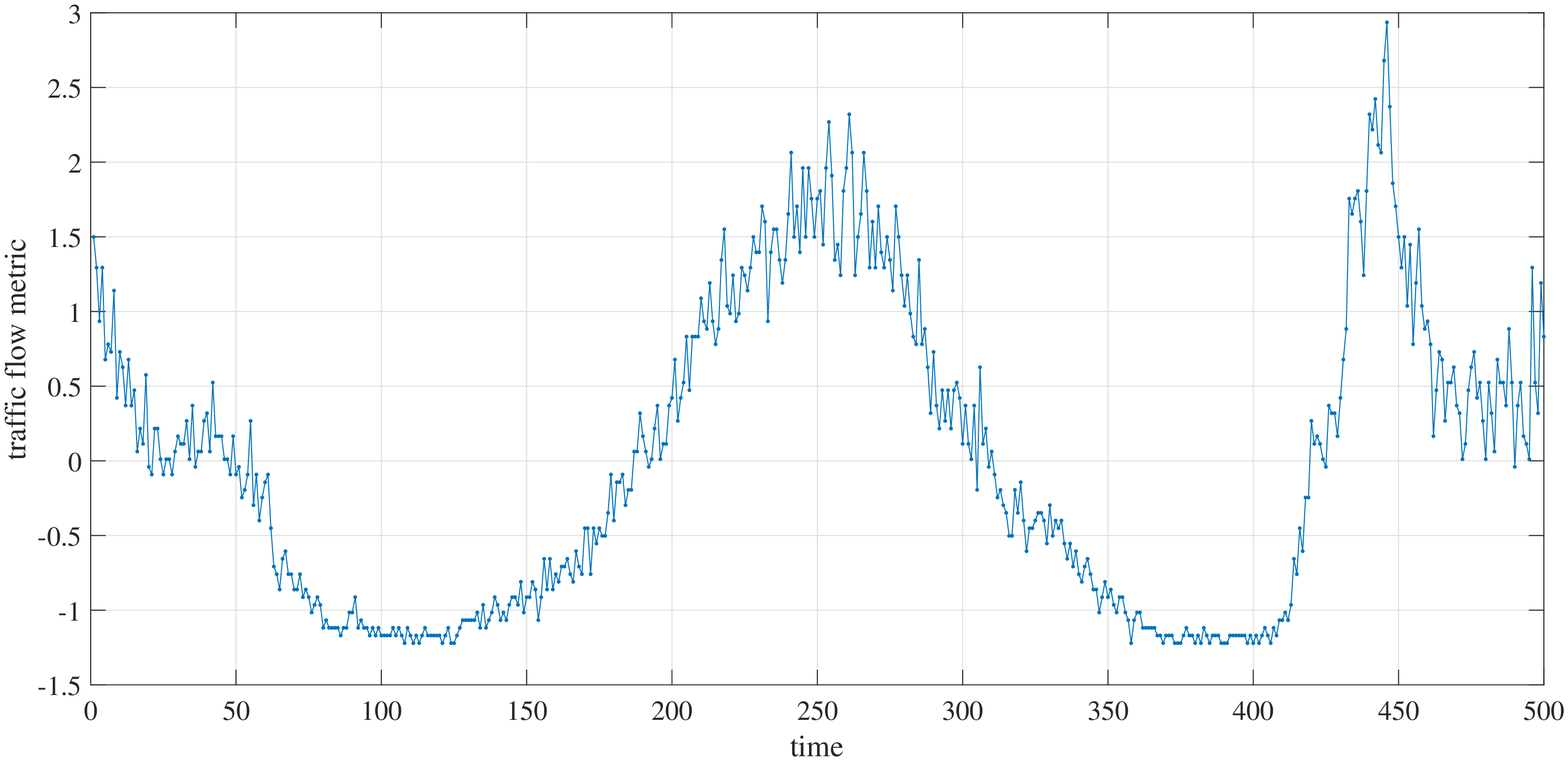}\\
\includegraphics[width=1.8in,height=0.8in]{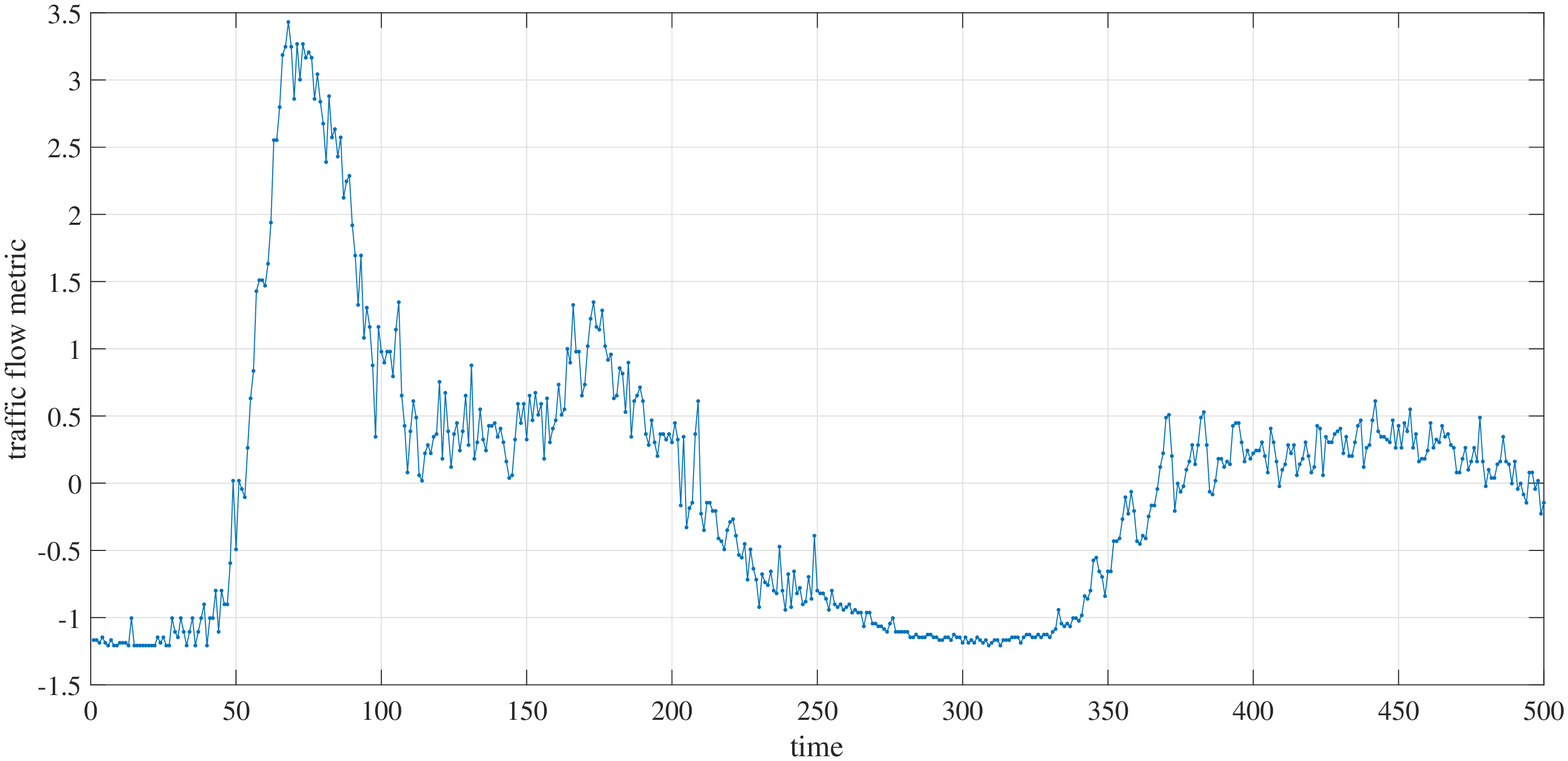}\includegraphics[width=1.8in,height=0.8in]{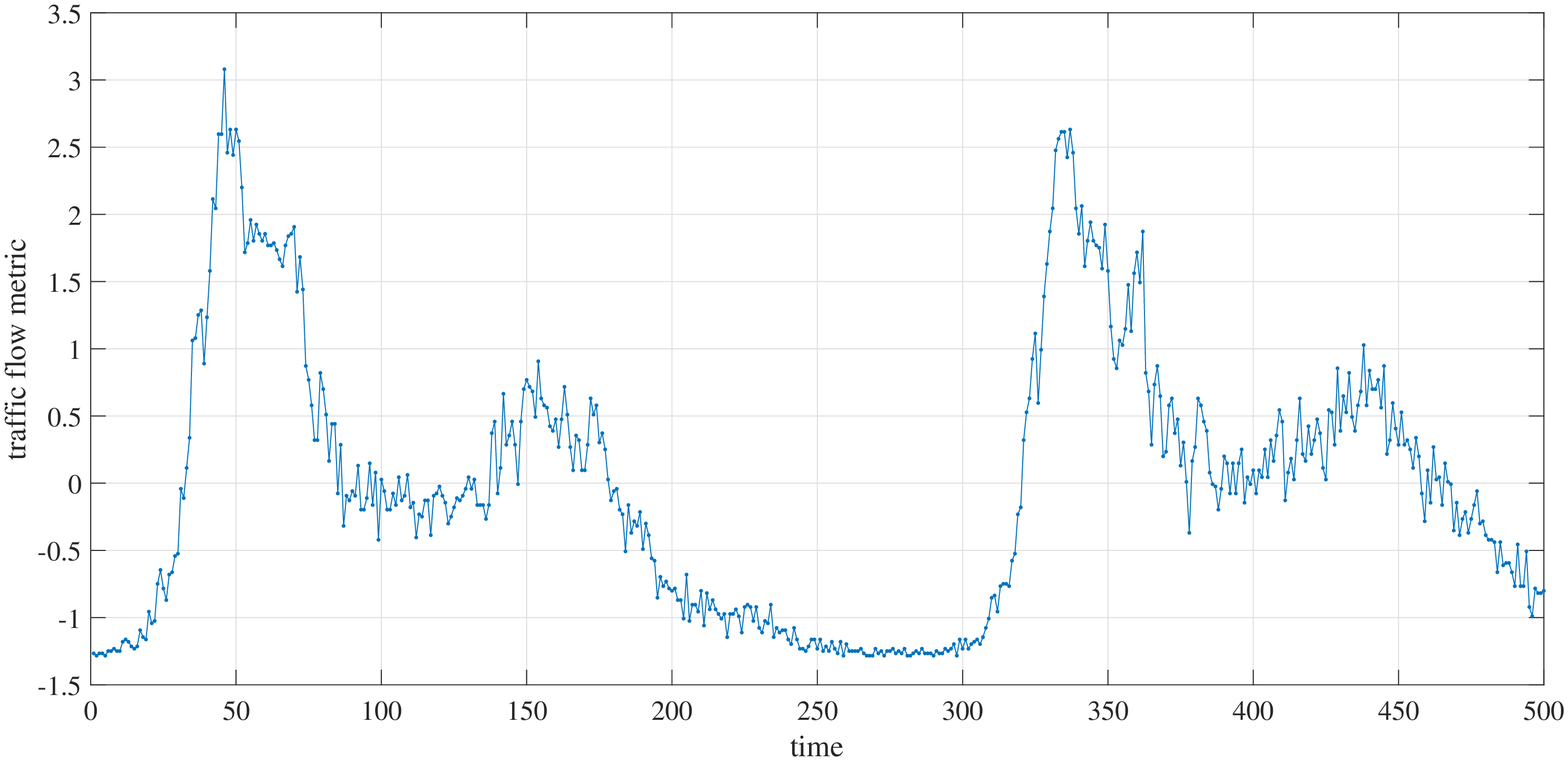}
\caption{Four segments of time-series from the PeMS dataset \cite{chen2001freeway}. They are treated as historical datasets that have been stored at 4 non-target edge servers. They are used by GPTDF-II for temporal data fusion.}\label{fig:dataset2}
\end{figure}

\begin{figure}[!htb]
\centering
\includegraphics[width=3.7in,height=1.8in]{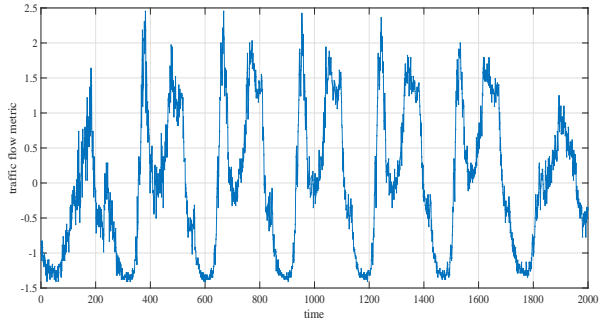}
\caption{One segment of time-series from the PeMS dataset \cite{chen2001freeway}, which is taken as the data that will be observed and processed at the target edge server}\label{fig:test_data}
\end{figure}

The performance metrics used for performance evaluation include the negative log likelihood (NLL),
the mean absolute error (MAE), and the mean square error (MSE). For each metric, the smaller is its value, the better the prediction
performance it stands for. The resulting prediction performance metrics are presented in Table \ref{Table:compare}. ``GPTDF-All" in
Table \ref{Table:compare} stands for the GPTDF method that fuses all 18 historical datasets in making predictions at the target edge
server. Both ``GPTDF-I" and ``GPTDF-II" only fuse 4 of the 18 historical datasets in making predictions at the target edge server.
The 4 datasets associated with ``GPTDF-I" are plotted in Fig.\ref{fig:dataset1}, and those used by ``GPTDF-II" are plotted in
Fig.\ref{fig:dataset2}. The observations that are processed at the target edge server, which we call the target dataset here, are
shown in Fig.\ref{fig:test_data}. ``GP" in Table \ref{Table:compare} represents the GP based prediction method that uses the first $N$ data points of the target dataset
for training the GP model. For ``GP-I", ``GP-II" and ``GP-III", $N$ takes values 50, 100 and 150, respectively.

As shown in Table \ref{Table:compare}, compared with the non-data-fusion based GP method, the biggest feature of our GPTDF method is
a zero-delay in making the predictions. This benefit comes from the temporal data fusion operation, which makes it unnecessary to gather
training data to train the model beforehand. In contrast, the working of the non-data-fusion based GP method requires
that a number of training data points be available beforehand for use in training the GP model.

It is shown that the best prediction performance is given by ``GPTDF-I", as it produces the most accurate prediction,
which is comparable with ``GP-III", while at a cost of zero-delay. In contrast, if ``GP-III" is adopted at the edge,
one has to bear the cost of the maximum delay. ``GPTDF-I" performs better than ``GPTDF-All".
It indicates that, for GPTDF, using more models is not certain to lead to better prediction performance.

\begin{figure}[!htb]
\centering
\includegraphics[width=3.7in,height=1.8in]{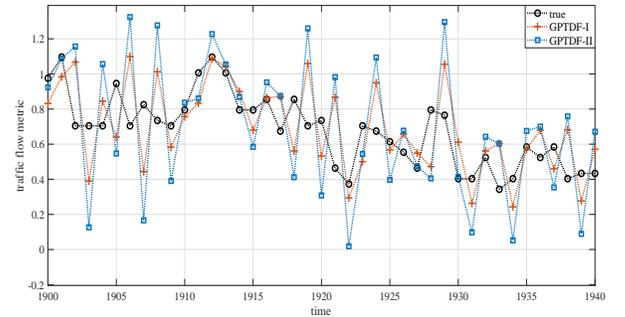}
\caption{Prediction performance comparison between GPTDF-I and GPTDF-II. See the text in Section \ref{sec:experiment} for definitions of GPTDF-I and GPTDF-II.}\label{fig:pred_compare}
\end{figure}

Despite that ``GPTDF-I" and ``GPTDF-II" fuse temporal feature information from the same number of historical datasets,
the difference in prediction performance between them is significant. This is again confirmed in Fig.\ref{fig:pred_compare}.
The GP model hyper-parameter values used in ``GPTDF-I" and ``GPTDF-II" are presented in Tables \ref{Table:hyper1}
and \ref{Table:hyper2}, respectively. As is shown, it is the difference in the hyper-parameter values
that leads to a significant difference in the prediction performance. In practice, how to select the best subset of
the model set that covers all available candidate models is a question to raise up. An easy solution is just to
use ``GPTDF-All", as it can produce moderate prediction accuracy at the cost of zero-delay.
\begin{table}[t]\centering\small
\caption{Hyper-parameter values of the candidate GP models employed by ``GPTDF-I"}
\begin{tabular}{c|c|c|c}
\hline %
  & $\sigma_l$ & $\sigma_f$ & $\sigma_n$ \\\hline
$\mathcal{M}_1$ & 2.0752  & 0.8215 & 0.1001 \\\hline
$\mathcal{M}_2$ & 2.4335 & 0.8069 & 0.1000 \\\hline
$\mathcal{M}_3$ & 2.2916 & 0.8096 & 0.1001 \\\hline
$\mathcal{M}_4$ & 2.1494 & 0.8206 & 0.1000 \\\hline
\end{tabular}
\label{Table:hyper1}
\end{table}
\begin{table}[t]\centering\small
\caption{Hyper-parameter values of the candidate GP models employed by ``GPTDF-II"}
\begin{tabular}{c|c|c|c}
\hline %
  & $\sigma_l$ & $\sigma_f$ & $\sigma_n$ \\\hline
$\mathcal{M}_1$ & 7.3899  & 0.7773 & 0.1000 \\\hline
$\mathcal{M}_2$ & 4.5846 & 0.7778 & 0.1007 \\\hline
$\mathcal{M}_3$ & 9.6141 & 0.7897 & 0.1001 \\\hline
$\mathcal{M}_4$ & 7.5284 & 0.8471 & 0.1003 \\\hline
\end{tabular}
\label{Table:hyper2}
\end{table}

As for the ``GP" method, it is shown in Table \ref{Table:compare} that, the more training data being used, the higher the prediction
accuracy, and the larger the delay in making the first prediction.
Note that since the ``GP" method uses the first $N$ data points of the target dataset to do model training before making predictions
for the follow-up observations, the delay for it making the first prediction is exactly $N$ time steps.
\section{Concluding Remarks}\label{sec:conclusion}
In this paper, for the first time, we proposed the concept of temporal data fusion at the edge. Our goal is to combine the strengths of edge computing and temporal data fusion by novel algorithm design. We focused on an application scenario, namely temporal data fusion assisted sequential online prediction at the edge, and proposed the GPTDF method, which inherits desirable properties of edge computing,
such as privacy-preserving, low latency between data capture and computation, and low bandwidth consumption.  We experimentally validated that the application of GPTDF can provide more timely and accurate predictions at the edge. In this way, we gave a proof-of-concept for temporal data fusion at the edge.

Currently, we only consider the fusion of homogeneous temporal data at the edge, while an interesting question is how to do the fusion of heterogeneous temporal data at the edge. The fusion mechanism may be application dependent. Besides, how to make use of the context information such as the edge servers' relative locations in the fusion process
is also worth further investigations.
\bibliographystyle{IEEEtran}
\bibliography{mybibfile}
\vspace{12pt}
\end{document}